# Micromagnetic understanding of the skyrmion Hall angle current dependence in perpendicular magnetized ferromagnets

Riccardo Tomasello[1], Anna Giordano[1], Stefano Chiappini[2], Roberto Zivieri[1], Giulio Siracusano[3], Vito Puliafito[4], Israa Medlej[1], Aurelio La Corte[3], Bruno Azzerboni[4], Mario Carpentieri[5], Zhongming Zeng[6], Giovanni Finocchio[1,2]

[1]*Dept. Mathematical and Computer Sciences, Physical Sciences and Earth Sciences, University of Messina, Messina, Italy*

[2]*Istituto Nazionale di Geofisica e Vulcanologia, Via di Vigna Murata 605, I-00143 Roma, Italy*

[3]*Department of Electric, Electronic and Computer Engineering, University of Catania, I-95125, Catania, Italy*

[4]*Dept. of Engineering, University of Messina, Messina, Italy*

[5]*Dept. Electrical and Information Engineering, Politecnico di Bari, Bari, Italy*

[6]*Key Laboratory of Nanodevices and Applications, Suzhou Institute of Nano-tech and Nano-bionics, Chinese Academy of Sciences, Ruoshui Road 398, Suzhou 215123, P. R. China*

**Abstract**

The understanding of the dynamical properties of skyrmion is a fundamental aspect for the realization of a competitive skyrmion based technology beyond CMOS. Most of the theoretical approaches are based on the approximation of a rigid skyrmion. However, thermal fluctuations can drive a continuous change of the skyrmion size via the excitation of thermal modes. Here, by taking advantage of the Hilbert-Huang transform, we demonstrate that at least two thermal modes can be excited which are non-stationary in time. In addition, one limit of the rigid skyrmion approximation is that this hypothesis does not allow for correctly describing the recent experimental evidence of skyrmion Hall angle dependence on the amplitude of the driving force, which is proportional to the injected current. In this work, we show that, in an ideal sample, the combined effect of field-like and damping-like torques on a breathing skyrmion can indeed give rise to such a current dependent skyrmion Hall angle. While here we design and control the breathing mode of the skyrmion, our results can be linked to the experiments by considering that the thermal fluctuations and/or disorder can excite the breathing mode. We also propose an experiment to validate our findings.

## I. INTRODUCTION

Magnetic skyrmions are localized swirling perturbations of the uniform magnetization texture [1–3] which can be stabilized by a sufficiently large Dzyaloshinskii–Moriya interaction (DMI) [4,5]. The DMI arises in systems with broken inversion symmetry, such as non-centrosymmetric bulk magnetic compounds (bulk DMI) [6–8], multi-layered heterostructures where a thin ferromagnet is in contact with a non-magnetic heavy metal with strong spin-orbit interaction (interfacial DMI - IDMI) [9–14], or in Heusler materials with tetragonal inverse structure (DMI in $D_{2d}$ structures) [15–17]. According to the kind of DMI, different types of skyrmions can be stabilized: Bloch skyrmion (vortex-like configuration) for bulk DMI materials, Néel skyrmion (hedgehog configuration) in interfacial DMI (IDMI) systems, and antiskyrmion in $D_{2d}$ structures.

Skyrmions can be very small (diameters of the order of nanometers), energetically stable, and topologically protected [2,3] because they are characterized by an integer winding number $S$ [18]. Owed to these features, together with the fact that they can be manipulated (nucleated, moved, and annihilated) by electrical currents [9,11,14,19], skyrmions are becoming attractive candidates to be used in low-power microelectronic applications, as building block of storage and programmable logic, as well as in alternative computational paradigms, such as probabilistic computing and skyrmion fabrics [20].

Recently, intense research efforts have led to the identification of multilayers heterostructures where Néel skyrmions are stable at room temperature and can be shifted by SOT [12–14]. To study the skyrmion stability, we have previously introduced a nonlinear ansatz [21] that can be used, together with proper scaling relations of the magnetic parameters, to analyze skyrmion stability and average size (diameter) as a function of the external field and temperature. In addition, it has been also shown that the thermal effects induce internal distortions of the skyrmion, which then loses its circular symmetry [21], expansion and shrinking of the skyrmion core (thermal breathing mode), and a thermal drift [22] (gyrotropic motion). The first result of this paper concerns the identification of the non-stationary origin of the skyrmion modes excited by thermal fluctuations.

Skyrmion motion can be driven by the spin-transfer torque (STT) [9,10,23] from an in-plane current flowing in the ferromagnets [24], or, more efficiently [9,10], by the spin-orbit torque (SOT) due to the spin-Hall effect (SHE) [25] and to the inverse spin-Galvanic effect [26], originating from a current flowing via a heavy metal with large spin-orbit coupling in contact with a ferromagnet. The skyrmion shifting is characterized by an in-plane angle with respect to the direction of the applied current, i.e. the skyrmion Hall angle (SHA) [27–29]. The control of the SHA is crucial for racetrack memory applications where skyrmions, coding the information, would be inevitably driven towards the sample edges where they could either bounce back or be annihilated. This aspect also limits the maximum applicable current and hence the maximum velocity achievable for the skyrmion. One strategy to overcome this issue is to suppress the SHA by balancing the Magnus force in two coupled skyrmions with opposite topological charge, resulting in zero net topological charge, in ferrimagnets [30], synthetic antiferromagnets [31,32], or antiferromagnets [33].

In an ideal ferromagnet/heavy metal bilayer, micromagnetic simulations and the theoretical approaches based on the Thiele's formalism including the damping-like torque (DLT) and the rigid skyrmion approximation, predict a constant SHA that is independent of the value of the applied current [10,27–29]. In particular, with our set of parameters [10], such a motion is characterized by a current-independent SHA $\phi_{SkH}$ equal to $\sim -90°$ with respect to the direction of a positive current (let us suppose positive $x$-axis). In other words, the skyrmion moves mainly along the negative $y$-direction with a small negative $x$-component of the velocity [10], and vice versa for negative currents.

On the contrary, recent experimental observations have shown a current dependence of the SHA. Jiang *et al.* [27] attributed such dependence to the presence of random pinning potentials in their materials. Basically, at low currents in the so-called creep regime, the skyrmion motion is strongly affected by pinning from defects and its SHA is current-dependent. Whereas, at high currents, skyrmion is characterized by a steady-flow regime where its SHA is independent of the current. Differently, Litzius *et al.* [28], claimed that the current dependence of the SHA results from the combination of the SHE-field-like torque (FLT) [34] and the internal deformations of the skyrmion. In order to add our contribution to this debate, we performed micromagnetic simulations where FLT and DLT are simultaneously applied to a breathing skyrmion. Our results show the qualitative picture that fundamentally explains the current dependent SHA. The paper is organized as follows. Section II presents the details of the micromagnetic framework. Section III shows the results of this work, with particular focus on the identification of the thermal breathing modes and on the control of the skyrmion Hall angle. Section IV is dedicated to the discussion of our results, where we also propose an experiment to quantify the effect of the FLT. The conclusions are reported in Section V.

## II. MICROMAGNETIC MODEL

We analyze a square sample where a thin ferromagnetic layer is in contact with a Platinum underlayer in order to obtain the IDMI and SHE. The micromagnetic study is performed by means of state-of-the-art processing tools and home-made micromagnetic solver GPMagnet [35,36] which numerically integrates the Landau-Lifshitz-Gilbert-Slonczewski equation by applying the time solver scheme Adams-Bashforth, where the SHE-FLT and SHE-DLT are taken into account:

$$\frac{d\mathbf{m}}{d\tau} = -(\mathbf{m} \times \mathbf{h}_{\text{eff}}) - \alpha\left(\mathbf{m} \times \frac{d\mathbf{m}}{d\tau}\right) - d_j\left[\mathbf{m} \times (\mathbf{m} \times (\hat{z} \times \mathbf{j}_{\text{HM}}))\right] - \nu d_j(\mathbf{m} \times (\hat{z} \times \mathbf{j}_{\text{HM}})) \quad (1)$$

where $\mathbf{m} = \mathbf{M}/M_s$ is the normalized magnetization of the ferromagnet, and $\tau = \gamma_0 M_s t$ is the dimensionless time, with $\gamma_0$ being the gyromagnetic ratio, and $M_s$ the saturation magnetization. $\mathbf{h}_{\text{eff}}$ is the normalized effective field, which includes the exchange, IDMI, magnetostatic, anisotropy and external fields. $\alpha$ is the Gilbert damping. $d_j = \frac{g\mu_B \theta_{SH}}{2\gamma_0 e M_S^2 t_{FM}}$ where $g$ is the Landé factor, $\mu_B$ is the Bohr magneton, $\theta_{SH}$ is the spin-Hall angle, $e$ is the electron charge, $t_{FM}$ is the thickness of the ferromagnetic layer. $\hat{z}$ is the unit vector along the out-of-plane direction, and $\mathbf{j}_{\text{HM}}$ is the electrical current density flowing into the Pt heavy metal which gives rise to the SHE. $\nu$ is a coefficient linking the magnitude of the FLT to the one of the DLT [34].

The lateral dimension of the square sample under investigation is 1.6 μm, while the ferromagnet thickness is $t_{FM}$=1 nm. The discretization cell used in the simulations is 2.0x2.0x1.0 nm$^3$. We introduce a Cartesian coordinate system with the *x*- and *y*- axes lying into the plane of the sample, whereas the *z*-axis is oriented along the out-of-plane direction. We consider typical parameters for CoFeB/Pt bilayers [10]: $M_S$=1000 kA/m, exchange constant $A$=20 pJ/m, perpendicular anisotropy constant $k_u$=0.80 MJ/m$^3$, $D$=2.0 mJ/m$^2$, $\alpha$=0.03, and $\theta_{SH}=0.1$. The current $j_{HM}$ flows in the *x*-direction, thus the spin-current inside the heavy metal is polarized along the *y*-direction. We apply an out-of-plane external field $H_z$=25 mT in all the simulations to have a metastable skyrmion [21] and to reduce the transient breathing mode. The SHE-FLT has the direction of the spin-polarization (*y*-axis). In order to deeply understand the role of the SHE-FLT and the physics underlying the SHA current dependence, we consider, in the first part of the paper, a constant FLT $|\nu d_j j_{HM} M_s| = |H_y| = 50$ mT, while, in the second part, we fix the DLT and we change the FLT.

For stochastic simulations, the thermal fluctuations are added to the deterministic effective magnetic field as a random term in each computational $\mathbf{h}_{\text{th}} = (\chi/M_S)\sqrt{2(\alpha K_B T / \mu_0 \gamma_0 \Delta V M_s \Delta t)}$, with $K_B$ being the Boltzmann constant, $\Delta V$ the volume of the computational cubic cell, $\Delta t$ the simulation time step, $T$ temperature of the sample, and $\chi$ a three-dimensional white Gaussian noise with zero mean and unit variance [37,38]. The noise is assumed to be uncorrelated for each computational cell.

# III. RESULTS

## A. Non-stationary behavior of thermal breathing modes

Thermal fluctuations can drive both a drift of the skyrmion, which has been also experimentally observed [39], and thermal modes, whose presence has been only shown by micromagnetic simulations [21] up to date. In this study, we have developed a technique that identifies such thermal modes from the time domain evolution of the skyrmion area as computed by micromagnetic simulations. For this purpose, we combine the Hilbert Huang transform [40–42] (HHT, see Supplementary note 1 [43]) and the wavelet transform [44,45]. The former aims at decomposing the signal into independent oscillatory components, also known as Intrinsic Mode Functions (IMF), in order to identify the time domain traces of the thermal modes. The latter studies the non-stationary behavior of those modes. We consider the data computed in our previous work [21]. In particular, the magenta curve in Fig. 1(a) represents an example of the time trace of the skyrmion area at $T$=300 K and $H_z$=25 mT (similar results are obtained for each combination of temperature and external field, where temperature ranges from 50 K to 300 K with steps of 50 K and the field is 0, 25 and 50 mT, see Supplemental Material [43], Movie 1). The skyrmion area is considered as the area of the region where the magnetization of the skyrmion core (negative $z$-component) is smaller than zero [21].

The HHT based computations show the existence of two IMFs that are correlated to the time domain trace of the skyrmion area. In Fig. 1(a), those two IMFs (IMF4 in blue and IMF5 in red) are represented as a function of time. Although the HHT is a powerful tool for analyzing complex datasets, many irrelevant IMFs may appear in the decomposition. A statistical significance of IMFs [46,47] uses a statistically-based threshold to discriminate between relevant and irrelevant IMFs. However, it has been shown how, for very noisy signals, both of these methods perform poorly [48]. Here, the criterion used for the classification of the IMFs in terms of significance is based on the evaluation of the normalized Cross-Correlations (CC) at lag of zero [49,50] between the signal of the skyrmion area, $x$, and each IMF component (a complete description of the method is provided in the Supplementary note 1 [43]). In detail, the HHT decomposes the signal, $x(t)$, into a finite number $N$ of different IMFs $y_i = y_i(t)$ (for $i = 1,…, N$), such that $x \cong \sum_{i=1}^{N} y_i$. For the $i^{th}$ IMF, the cross-correlation with $x$, $\hat{R}_{x,y_i}$ can be defined as [50]:

$$\hat{R}_{x,y_i}[m] = \begin{cases} \sum_{n=0}^{L-m-1} x[n+m] y_i^*[n] & m > 0 \\ \hat{R}_{y_i,x}^*[-m] & m < 0 \end{cases} \quad (2)$$

where $y_i^*$ is the complex conjugate of $y_i$. The value $CC_i = \dfrac{\hat{R}_{x,y_i}[0]}{\|x\|\cdot\|y_i\|}$ represents the normalized cross-correlation at lag of zero ($-1 \leq CC_i \leq 1$). The correlation is computed according to Eq. (2) and the results of the computation for all the IMFs are reported in Table I for the case under investigation ($H_z$=25 mT, T=300 K). The relevance of IMFs is established by evaluating the corresponding CC values and peak frequency values ($f_{peak}$). For example, for the case investigated in Fig. 1, the most relevant IMFs are IMF4 and IMF5, although there are other modes with high CC, but with relatively low frequency fluctuations (such as the sub-GHz modes indicated with IMF11 and IMF15), which are far from the frequencies obtained by deterministic micromagnetic simulations (see below).

Table I. Results of HHT for the skyrmion area signal as investigated in Fig.1. Cross-correlation (CC) of corresponding IMFs and peak frequency values ($f_{peak}$), respectively.

| IMF number | |CC| | $f_{peak}$ (GHz) |
|---|---|---|
| IMF1 | 0.029 | 58.86 |
| IMF2 | 0.020 | 27.56 |
| IMF3 | 0.046 | 7.41 |
| **IMF4** | **0.128** | **5.33** |
| **IMF5** | **0.154** | **3.54** |
| IMF6 | 0.117 | 2.04 |
| IMF7 | 0.101 | 1.07 |
| IMF8 | 0.077 | 0.115 |
| IMF9 | 0.041 | 0.142 |
| IMF10 | 0.0911 | 0.121 |
| IMF11 | 0.171 | 0.079 |
| IMF12 | 0.083 | 0.031 |
| IMF13 | 0.061 | 0.021 |
| IMF14 | 0.039 | 0.010 |
| IMF15 | 0.331 | 0.005 |

In addition to the Fast Fourier Transform that shows a main frequency component of 5.45 GHz for the IMF4 and of 3.54 GHz for the IMF5, we perform a wavelet analysis (see Supplementary note 2 [43]). Figs. 1(b) and (c) show the wavelet scalogram [51] that clearly brings the conclusion of the non-stationary feature of those two thermal modes.

To show that those excitations are linked to the skyrmion breathing, we carried out deterministic micromagnetic simulations by using the value of the physical parameters as calculated by scaling relations [21,52] (see Supplementary note 3 [43] and Supplemental Material [43], Movie 2). We observe that the FMR frequency of the skyrmion breathing mode lies inside the frequency range identified by the most significant IMFs' frequencies. For instance, for the case shown in Fig. 1 ($H_z$=25 mT, $T$=300 K), the FMR frequency is 5.0 GHz and it is comprised between 3.54 GHz (IMF5) and 5.45 GHz (IMF4). This achievement, that is the first main result of this paper, is robust for each field-temperature combination, as can be observed in Fig. 2 that displays the FMR frequencies (symbols in the Figure) and the thermal modes frequencies (dashes in the Figure) as a function of temperature (50-300 K) at three different fields (0, 25 and 50 mT).

## B. Study of the skyrmion Hall angle

The computations discussed in the previous paragraph show that thermal fluctuations excite non-stationary breathing modes. It is also well known that disorder physical parameters can give rise to a continuous change of the skyrmion size [53], hence we conclude that the non-stationary breathing mode should have a role in the explanation of the current dependence SHA found in the experiments [27,28]. In addition, it has been already shown that the combination of an in-plane field and a breathing mode induces a Bloch skyrmion shift [54]. With this in mind, we have designed a numerical experiment to investigate the dynamical properties of a breathing Néel skyrmion driven by the FLT and DLT. In order to excite the skyrmion breathing mode by means of an ac perpendicular spin-polarized current (see sketch in Fig. 3(a)) to resemble the effect of thermal fluctuations and/or disordered physical parameters, we add to Eq. (1) the following STT term [55]:

$$-\frac{g\mu_B j_P}{\gamma_0 e M_S^2 t_{FM}} \varepsilon_{MTJ}(\mathbf{m},\mathbf{m_p}) \left[\mathbf{m}\times\left(\mathbf{m}\times\mathbf{m_p}\right)\right] \quad (3)$$

where $j_P = J_{AMP}\sin(2\pi ft)$ is the ac current density, with $J_{AMP}$ and $f$ being its amplitude and frequency, respectively. $\varepsilon_{MTJ}(\mathbf{m},\mathbf{m_p}) = \frac{2\eta}{\left[1+\eta^2\left(\mathbf{m}\cdot\mathbf{m_p}\right)\right]}$ is the polarization function [56,57], where $\eta = 0.66$ is the spin polarization factor, and $\mathbf{m_p}$ is the magnetization of the polarizer, which is considered fixed along the out-of-plane direction, thus generating a perpendicularly polarized spin-current.

For this study, we consider the same parameters already used in Ref. [10] for a direct comparison. The features of the skyrmion dynamics are described in terms of SHA and *x*- and *y*- components of the skyrmion velocity. To this aim, we introduce a cartesian coordinate reference system, as indicated in the inset of Fig. 5(a), where also the SHA is represented.

### 1. *Characterization of the breathing modes*

The skyrmion dynamics in presence of a persistent breathing mode are studied at $T$=0 K and $H_z$=25 mT, in order to obtain some simple and fundamental achievements. Fig. 3(b) shows the FMR response of the skyrmion, where the peak-to-peak amplitude of $<m_z>$ is plotted as a function of the ac current frequency $f$. For the range of $J_{AMP}$ plotted, the response is linear with a FMR frequency equal to 2.9 GHz, in qualitative agreement with previous studies [55,58]. However, the presence of $H_z$ introduces an upper threshold for the applied current (in our case is 2.0 MA/cm$^2$) over which the skyrmion is annihilated for frequencies near the FMR one. This occurs because, during half period of the current, $H_z$ favors the skyrmion shrinking which becomes critical over a certain value of the current amplitude.

For the current-driven dynamics of the skyrmion, to avoid skyrmion annihilation, we consider ac currents with frequencies larger than the FMR one, i.e. $f \geq 3.0\,\text{GHz}$, and we will focus on two cases: small breathing mode, where the perpendicular current amplitude $J_{AMP}$ is 2 MA/cm$^2$, and large breathing mode, with $J_{AMP}$ equal to 6 MA/cm$^2$ (note that far from the resonant frequency the skyrmion does not annihilate at this current density).

### 2. *Effect of the field-like torque*

When considering a rigid skyrmion, its motion driven by the DLT is not affected by the FLT (see Supplemental Material [43], Movie3). In particular, the role of the FLT is to elongate the skyrmion along the direction of the in-plane field [54,59] leading to a noncircular skyrmion, without modifying the skyrmion velocity and SHA, but only the maximum applicable current. This result is in agreement with previous works on Bloch skyrmions [59]. Figure 4 displays such a deformation along the field direction, where the main panel shows the spatial-averaged *y*-component of the skyrmion magnetization $<m_{y\_sk}>$ as a function of the in-plane field. As expected, $<m_{y\_sk}>$ is almost zero at zero field because the skyrmion is symmetric and circular, while it increases linearly with the field. This result is confirmed by the spatial distribution of the skyrmion magnetization (insets in Fig. 4), where we can observe how the skyrmion becomes non-circular as the field increases.

### 3. *Effect of the breathing mode*

Let us now consider a breathing skyrmion. Its motion driven by the DLT in term of SHA does not differs from the rigid one, because the periodic change of the skyrmion radios has a null effect on average (not shown, See Supplemental Material [43], Movie3). On the contrary, the FLT gives rise to a skyrmion shift along the x-direction (same direction of the electrical current) because the in-plane field breaks the skyrmion symmetry, as already predicted for Bloch skyrmions [54]. Therefore, the breathing skyrmion undergoes two antagonistic effects: from one side, the DLT moves it mainly along the y-direction, from the other side, the FLT moves it along the x-direction. This aspect will allow us to control the SHA, as it will be shown in the next paragraph.

### 4. *Dynamics driven by SHE-DLT and SHE-FLT in presence of breathing mode*

In this paragraph, we show the results obtained when SHE-DLT, SHE-FLT act simultaneously on a breathing skyrmion.

We fix the current $j_{HM}$ flowing into the heavy metal to -5 MA/cm$^2$ and we use two values of FLT that correspond to $H_y=\pm 50$ mT. Figs. 5(a)-(c) show the results for small breathing. Although there is a slight variation of $\phi_{SkH}$, $v_x$, and $v_y$, this can be considered negligible. On the contrary, when the breathing becomes larger (Figs. 5(d)-(f)), we observe that the *x*-component of the velocity is negative for all the frequencies for negative values of the in-plane field. In other words, as expected, the FLT affects mainly $v_x$, promoting a skyrmion displacement along the positive (negative) x-direction for positive (negative) values of the field, while the DLT influences the y-component of the velocity. It is important to notice that, in Fig. 1d, for $H_y$=-50 mT, $\phi_{SkH}$ is computed with the respect to the negative *x*-axis. In addition, the small change of the $\phi_{SkH}$ as a function of frequency is still negligible. In the rest of the paper, we will focus on a frequency of 8 GHz to speed up the computations, however simulations performed at different frequencies do not exhibit any qualitative difference.

We now study the skyrmion dynamics by changing the value of $j_{HM}$ (SHE-DLT), at a fixed $H_y=\pm 50$ mT (SHE-FLT). It is interesting to observe in Fig. 6 that also for $j_{HM}$=0 MA/cm$^2$ we obtain the skyrmion motion, but with a major component along the positive (negative) x-direction if the field is positive (negative). In other words, the combination of only FLT and breathing mode promotes the skyrmion shift in the same direction of the electrical current. When we increase the magnitude of $j_{HM}$, the DLT, which drives the skyrmion mainly in the direction perpendicular to the electrical current, leads $\phi_{SkH}$ to be almost 90° from very low currents in the case of small breathing ($j_{HM}$>1 MA/cm$^2$ in Fig. 6(a)). The x-component of the velocity (Fig. 6(b)) slightly changes with current. This is due to the fact that it is of the same order of the one induced by the DLT. In particular, a positive (negative) current introduces a $v_x$ of the same sign as the one due to the FLT for negative (positive) fields, hence the magnitude of $v_x$ increases. The y-component of the velocity is not affected by the FLT (Fig. 6(c)) because the blue and black curves coincide.

If the breathing is larger, $\phi_{SkH}$ slowly increases with current and starts to saturate to 76° for currents larger than 5 MA/cm$^2$ (Fig. 6(d)). This behavior is in qualitative agreement with the experimental evidences [27,28] (see, for instance, Fig. 3c of Ref. [27]). In addition, it is more evident that the combination of FLT and breathing mode introduces a smaller y-component of the velocity. In fact, at zero $j_{HM}$, $\phi_{SkH}$ is about 30°. The larger breathing mode enhances also the effect of the FLT, i.e. $v_x$ is almost constant with current, meaning that the x-axis skyrmion motion is dominated by the FLT (Fig. 6(e)). Owed to the smaller $v_y$ induced by the FLT, also the total $v_y$ is affected, in fact the two curves are no longer overlapped (Fig. 6(f)). In all the cases, we observe a symmetric skyrmion dynamics with respect to the sign of the current.

Finally, we study the skyrmion dynamics by changing the value of $H_y$ (SHE-FLT), when $j_{HM}=\pm 5$ MA/cm$^2$. When the breathing is small (Figs. 7(a)-(c)), the FLT slightly affects only the x-component of the velocity, while $\phi_{SkH}$ and $v_y$ keep constant with fields. In particular, the magnitude of $v_x$ increases with positive (negative) fields when the current is negative (positive), as explained above for Fig. 6(b).

When the breathing mode becomes larger (Figs. 7(d)-(f)), the FLT has a dominant effect. The magnitude of $\phi_{SkH}$ increases (decreases) for negative (positive) values of $H_y$, while $v_x$ is almost independent of the DLT (Fig. 7(e)). In addition, Fig. 7(f) shows that the y-component of the velocity, which is typically due to the DLT, is also affected by the FLT, that leads to a larger magnitude of $v_y$ when the it has the same direction as the DLT (same sing of $H_y$ and $j_{HM}$). Again, these achievements demonstrate that an appropriate combination of SHE-DLT, SHE-FLT and breathing mode allows one to control the direction of the skyrmion motion.

## IV. DISCUSSION

These results clearly show that the key aspect, which gives rise to the experimental current-dependence of the SHA [27,28], is the combination of the breathing mode (due to thermal fluctuations and/or internal defects) and a sufficiently large SHE-FLT. For the sake of clarity, we have included Table II that summarizes the properties of the skyrmion motion by combining all of those ingredients. As it can be observed, the slope of the region where $\phi_{SkH}$ depends on current is smaller (larger current region) as the amplitude of the breathing mode increases. In detail, if the breathing is small (results for $J_{AMP}$=2 MA/cm$^2$), the SHA saturates for $|j_{HM}|$>2 MA/cm$^2$. When the breathing is large (results for $J_{AMP}$=6 MA/cm$^2$), the SHA is still not completely saturated up to $|j_{HM}|$>10 MA/cm$^2$. In other words, the larger is the breathing mode the larger is its effect on the skyrmion dynamics and therefore on the range of current dependence of the SHA. If we want to link this result to an experiment, it means that a more pronounced breathing, i.e. larger disorder and/or temperature, will lead to a larger region where the $\phi_{SkH}$ depends on current. To confirm those conclusions and resolve the debate discussed in the introduction, it would be sufficient to characterize $\phi_{SkH}$ as a function of an in-plane field for different currents. When the in-plane field is parallel to the field-like torque, $\phi_{SkH}$ current dependence should be more evident (smaller slope, Fig. 6(d) for example), as compared to the case when the in-plane field partially or totally compensates the FLT, where such a dependence should be almost absent.

Table II. Summary of the directions of motion of skyrmion under the effect of the different sources. X indicates which source of motion is active. FLT and DLT are considered for positive field and current. [1] See Supplemental Material [43], Movie 3. [2] See Supplemental Material [43], Movie 4.

| Breathing mode | FLT | DLT | X direction | Y direction |
|---|---|---|---|---|
| X | | | no motion | no motion |
| | X | | no motion | no motion |
| | | X | negative (small)[1] | negative |
| X | X | | positive | negative (small) |
| X | | X | negative (small)[1] | negative |
| | X | X | negative (small)[1] | negative |
| X | X | X | negative/positive depending on DLT and FLT[2] | negative |

## V. SUMMARY AND CONCLUSIONS

In summary, we have performed a numerical experiment based on micromagnetic simulations to understand the effect of SHE-FLT and DLT on a non-rigid (breathing) skyrmion. We have preliminarily demonstrated that the breathing of the skyrmion can be originated by thermal fluctuations. We have also shown that the SHE-FLT and DLT promote two antagonist skyrmion motions: the former occurs mainly along the *x*-axis, while the latter occurs mainly along the *y*-direction. Therefore, the combination of the breathing mode and a sufficiently large SHE-FLT gives rise to the current dependence of the SHA, as observed in the experiments [27,28]. Our result could be confirmed by an experiment, where the SHA should be characterized as a function of an in-plane field for different currents. According to the amplitude of the range where the SHA exhibits the current dependence, it would be possible to estimate the effect of the SHE-FLT.

## VI. ACKNOWLEDGEMENTS

This work was supported by the executive programme of scientific and technological cooperation between Italy and China for the years 2016–2018 (2016YFE0104100) title "Nanoscale broadband spin-transfer-torque microwave detector" funded by Ministero degli Affari Esteri e della Cooperazione Internazionale (code CN16GR09).

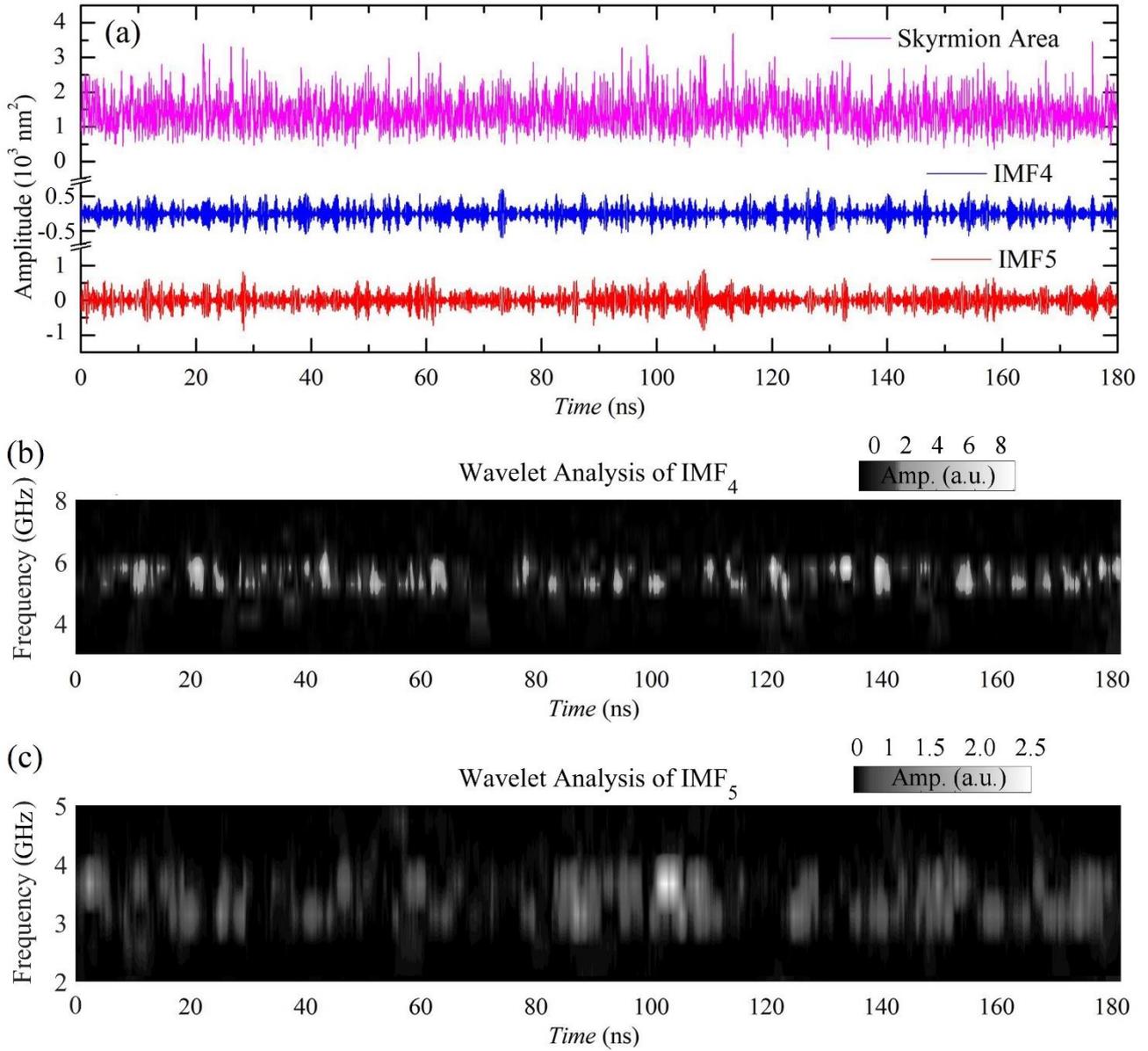

Figure 1: (a) Time evolution of the skyrmion area (magenta curve) when $H_z$=25 mT and $T$=300 K and corresponding most significant IMFs (blue and red curves), as obtained by HHT analysis. (b) and (c) Wavelet results of IMF4 and IMF5, respectively, where the color scale indicates the amplitude of the signal.

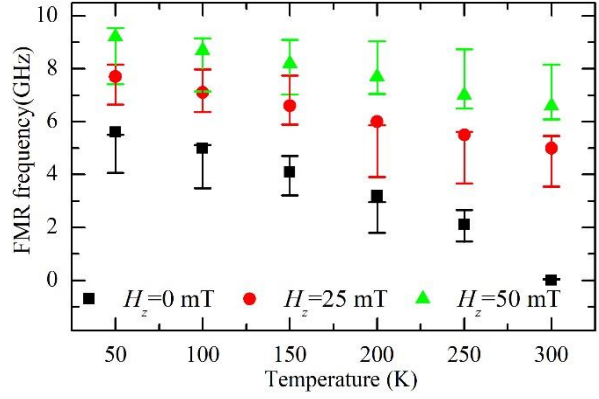

Figure 2: FMR frequency as a function of temperature and external field, where the symbols represent the results as obtained by micromagnetic simulations by using scaling relations of the physical parameters, while the two dashes indicate the frequency of the two most significant IMFs.

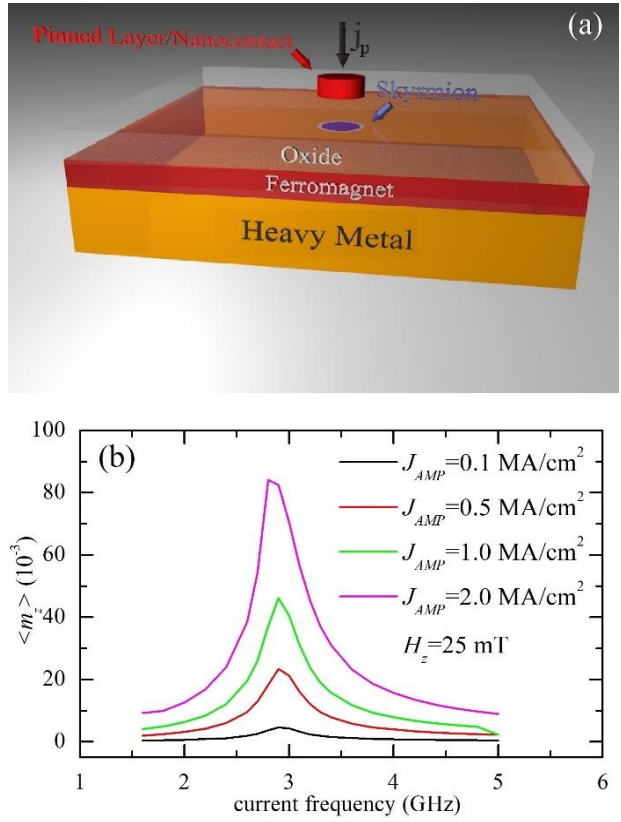

Figure 3: (a) sketch of the device under investigation to study the FMR response of the skyrmion, where the ac perpendicular current is locally injected via a nanocontact. (b) FMR frequency response driven by the ac perpendicular current when $H_z$=25 mT and $T$=0 K, for different values of $J_{AMP}$ as indicated in the legend.

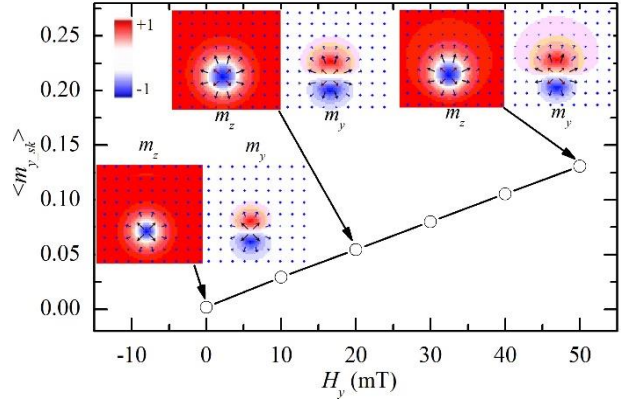

Figure 4: average *y*-component of the magnetization of the skyrmion as a function of the in-plane field. Inset: spatial distribution of the magnetization displaying the skyrmion for $H_y$=0 mT, 20 mT and 50 mT. The left panel represents the z-component of the magnetization, while the right panel represents the y-component. The same color scale for both components is also indicated.

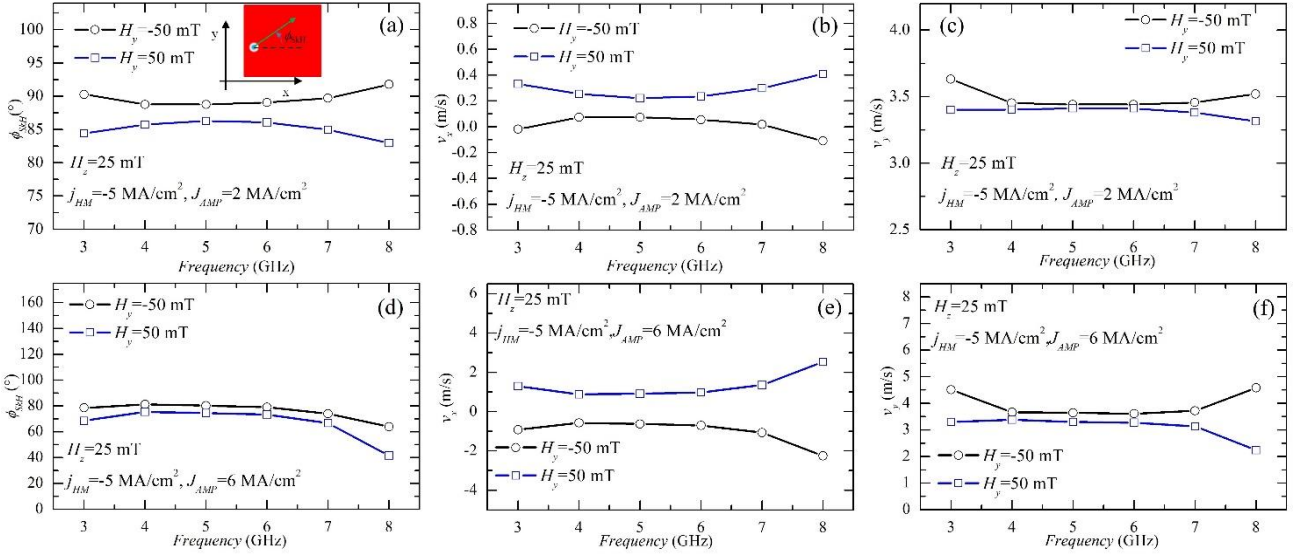

Figure 5: Skyrmion dynamics ($\phi_{SkH}$, $v_x$, and $v_y$) as a function of the breathing mode frequency for $H_y$=50 mT (blue curve with squares) and -50 mT (black curve with circles), when $H_z$=25 mT, $j_{HM}$=-5 MA/cm$^2$, and (a), (b), and (c) $J_{AMP}$=2 MA/cm$^2$, and (d), (e) and (f) $J_{AMP}$=6 MA/cm$^2$. In (a), $\phi_{SkH}$ is calculated with the respect to the positive *x*-axis. In (d), $\phi_{SkH}$ is calculated with the respect to the negative *x*-axis when $H_y$=-50 mT.

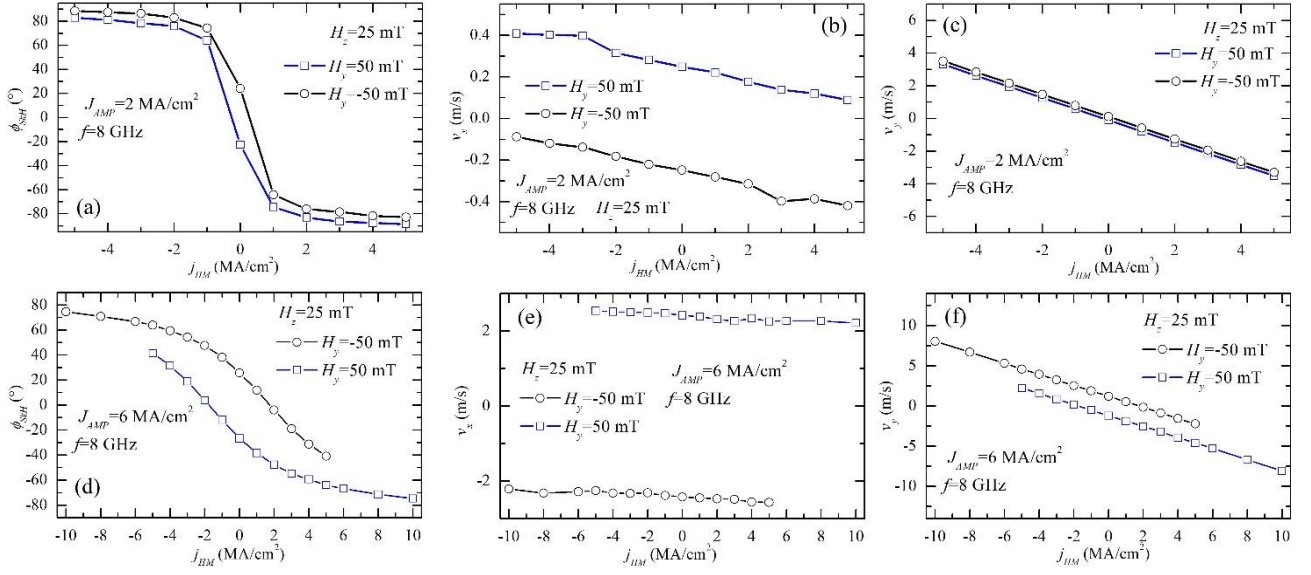

Figure 6: Skyrmion dynamics as a function of SHE-DLT, where $\phi_{SkH}$, $v_x$, and $v_y$, are represented with regards to the in-plane current $j_{HM}$, for $H_y$=50 mT (blue curve with squares) and -50 mT (black curve with circles), when $H_z$=25 mT, $f$=8 GHz, and (a), (b) and (c) $J_{AMP}$=2 MA/cm$^2$, (d), (e), and (f) $J_{AMP}$=6 MA/cm$^2$. In (a) and (d), $\phi_{SkH}$ is calculated with the respect to the negative $x$-axis when $H_y$=-50 mT.

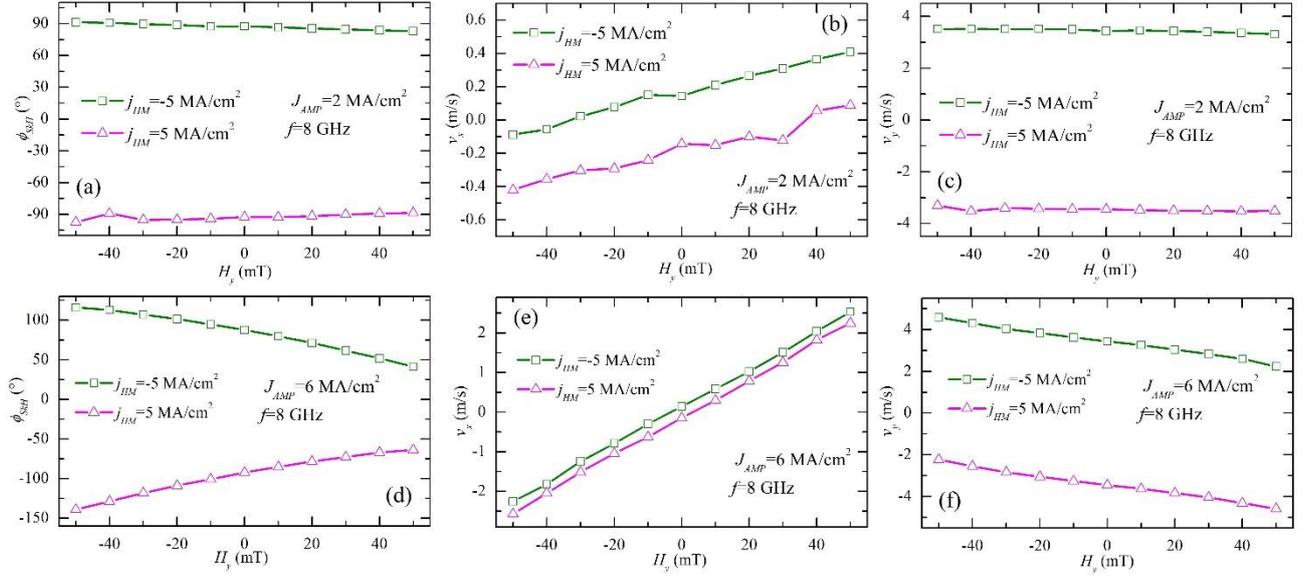

Figure 7: skyrmion dynamics as a function of the SHE-FLT, where $\phi_{SkH}$, $v_x$, and $v_y$, are plotted with regards to the in-plane field $H_y$, for $j_{HM}$=5 MA/cm$^2$ (olive curve with squares) and -5 MA/cm$^2$ (magenta curve with triangles), when $H_z$=25 mT, $f$=8 GHz, and–(a) - (c) $J_{AMP}$=2 MA/cm$^2$, (d) – (f) $J_{AMP}$=6 MA/cm$^2$. In (a) and (d), $\phi_{SkH}$ is calculated with the respect to the positive $x$-axis.

# SUPPLEMENTARY INFORMATION

## Supplementary note 1

The Hilbert-Huang Transform (HHT) is a recently developed method which has proved to be a useful tool for studying the nonstationary and nonlinear behavior of time series [1,2] with a lower computational cost as compared to Principal Component Analysis (PCA) and clusterization, or machine learning algorithms [3]. By means of this technique, complex sets of nonlinear and nonstationary data can be decomposed into a finite collection of individual characteristic oscillatory modes, named IMF. Since accurate filtering capabilities are the key ingredient in the study of skyrmion area, HHT is considered more appropriate for the full investigation of both nonstationary and nonlinear components of source signal than other techniques known in literature, especially for its better resolution in time and frequency domains [2].

HHT is a combination of Hilbert Transform (HT) operator and Empirical Mode Decomposition (EMD). EMD is an adaptive and efficient method applied to decompose nonstationary and nonlinear signal, which has been further enhanced through ensemble averaging method using signal added by white noise and made its averaging value as the real final output results [4]. In this method called Ensemble Empirical Mode Decomposition (EEMD), firstly, the ensemble datasets are generated by adding the original signals and white noise with amplitude. Then, EMD is performed on each ensemble dataset until the process reaches the last number of ensemble. The final value is obtained by averaging consecutive components resulted from the ensemble process. The goal of EEMD method is for obtaining better spectral separation at each oscillatory mode of the IMFs. Specifically, HHT with EEMD is the method we used for the further analysis in this manuscript.

The initial attempt at using HHT as a denoising tool emerged from the need to know whether a specific IMF contains or not useful information. Thus, significance IMF test procedures were recently developed [5] based on the statistical analysis of modes resulted from the decomposition of signals. Here, the classification of the EMD modes is based on the evaluation of the normalized Cross-Correlations at lag of zero (CC) between the signal and each Intrinsic Mode Function (IMF) and a threshold value $CC_{TH}$ to identify statistically relevant components. In detail, the HHT decomposes the signal, $x \equiv s_{k,z}(t)$, into $N$ different IMFs $y_i$ (for $i = 1,\ldots, N$), where $x \cong \sum_{i=1}^{N} y_i$. For the $i^{th}$ IMF, the cross-correlation with $x$, $\hat{R}_{x,y_i}$ can be defined as:

$$\hat{R}_{x,y_i}[m] = \begin{cases} \sum_{n=0}^{L-m-1} x[n+m] y_i^*[n] & m > 0 \\ \hat{R}_{y_i,x}^*[-m] & m < 0 \end{cases} \qquad (4)$$

where $y_i^*$ is the complex conjugate of $y_i$. The value $CC_i = \frac{\hat{R}_{x,y_i}[0]}{\|x\| \cdot \|y_i\|}$ represents the normalized cross-correlation at lag of zero ($-1 \leq CC_i \leq 1$). In our work, we investigated the independent oscillatory components of each input signal having the most relevant CC and whose frequency peak was in the range of interested of the physical phenomena under investigation.

**Supplementary note 2**

By performing the Fourier transform with a window time it can be observed the existence of a noisy Fourier spectrum which suggests the presence of multiple oscillatory components in the source signal. This result shows the intrinsic non-periodic behavior of this signal revealing also the presence of those two independent non-transient modes at nanosecond scale. From computational point of view, it is important to find out a tool which systematically gives information about the time localization of the excited modes. We use a wavelet-based analysis (the wavelet is the natural generalization of the Windowed Fourier transform) and differently by other approaches, we systematically identify the scale set directly from the power spectrum related to $r(t)$, which is the signal representing the skyrmion area as a function of time. The use of a wavelet analysis allows to characterize a signal in the time-frequency space to study the non-stationary behavior. The continuous wavelet transform (CWT) of the function $r(t)$ is a linear transform $W_r(u,s)$ given by [6]:

$$W_r(u,s) = \frac{1}{\sqrt{s}} \int_{-\infty}^{+\infty} r(t) \psi\left(\frac{t-u}{s}\right) dt \qquad (1)$$

Being $t$ the time, $s$ and $u$ the scale and translation parameters of the mother wavelet $\psi(t)$, which defines the wavelet family function as $\psi_{u,s} = \frac{1}{\sqrt{s}} \psi\left(\frac{t-u}{s}\right)$. In our study, in order to characterize both amplitude and phase of time-domain magnetoresistive signal $r(t)$ we use the complex Morlet wavelet family function as $\psi_{u,s} = \frac{1}{\sqrt{2\pi f_B}} e^{j2\pi f_C \left(\frac{t-u}{s}\right)} e^{-\frac{\left(\frac{t-u}{s}\right)^2}{f_B}}$, where $f_C$ and $f_B$ are two specific components, namely wavelet center frequency and bandwidth parameter, respectively (for a complete review of wavelet theory see Refs. [7,8]). Particularly, the $f_B$ term can rule the band of the Fourier spectrum of the complex Morlet function giving narrowed band as it increases, consequently this parameter is correlated to the frequencies we have to analyze independently of each other. In addition, from practical reasons $f_B$ and $f_C$ have to be large enough to made the mean of $\psi(t)$ arbitrarily small [9]. Our results suggest that the continue wavelet transform $W_r(u,s)$ shows better statistical performance than any other time-frequency analysis methods used to analyze of the simulated signals. Given a fixed dimension $N$ of a scale set $\{s_i\}_{i=1,...,N}$ for the wavelet family, the y-axis is divided in $N+2$ points (a complete description of the nonlinear decomposition method is provided in Ref. [10]) and for each point can be obtained a frequency $f_i$ and a scale $s_i$. The wavelet scalogram (WS) is used to show the time-frequency characterization of the signal, furthermore the integral of the signal over the time can be correlated to the Fourier spectrum of that signal directly for a fixed scale parameter [11]. Figures 1(b) and (c) in the main text show the WS (arb. units) as computed with the following parameters $N=16$, $f_B = 3000$, and $f_C = 1$.

## Supplementary note 3

The thermal effects can be included into the micromagnetic framework by setting scaled values of the macroscopic parameters. In particular, we calculate the temperature dependence of uniaxial magnetic anisotropy $K_u(T) = K_u(0)m(T)^\gamma$, exchange $A(m) = A(0)m^\alpha$ and DMI $D(T) = D(0)m(T)^\beta$, where $m(T) = M_s(T)/M_s(0)$ is the reduced magnetization, and $M_s(0)$, $K_u(0)$, $A(0)$, and $D(0)$ are the values of saturation magnetization, perpendicular anisotropy, exchange and DMI constants at zero temperature. The values of the scaling exponents are taken from Ref. [12]. In this way, we are able to analyze the effect of thermal fluctuations by performing deterministic micromagnetic simulations without introducing noise into the system.

The aforementioned approach allows us to study the skyrmion resonant response by locally excite the skyrmion with a spin-polarized current [13]. Specifically, the injection of a localized perpendicularly polarized current excites a steady breathing mode of the skyrmion, hence inducing a continuous change of the *z*-component of the spatial-averaged magnetization <$m_{z\_uc}$> under the nanocontact. The injection of a localized current also avoids the influence of the boundaries as well as of the background magnetization on the skyrmion response. Therefore, we apply a current with amplitude $J_{AMP}$=2 MA/cm² via a nanocontact with diameter 140 nm and sweep its frequency from 10 MHz to 10 GHz to find the ferromagnetic resonance (FMR) frequency for each combination of temperature and field.

Figure S1 shows some examples of FMR response curve for $H_z$=25 mT. The FMR response is linear when the skyrmion is metastable [13,14], with the FMR frequency increasing with field and decreasing with temperature.

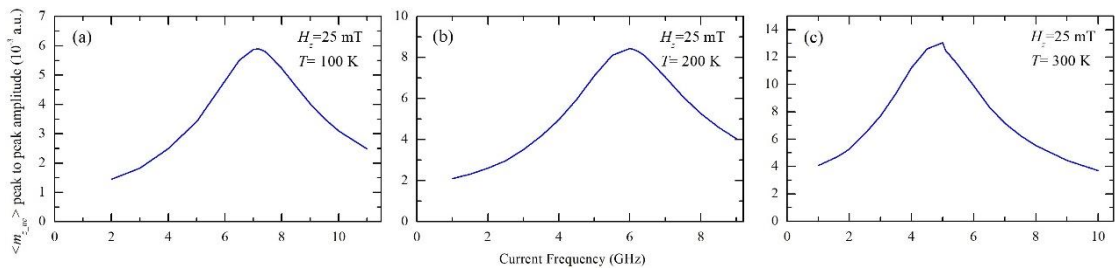

Figure S1: Resonant response of the skyrmion for $H_z$=25 mT when (a) *T*=100 K, (b) *T*=200 K and (c) *T*=300 K. In all the figures, the peak to peak amplitude of the z-component of the magnetization under the nanocontact is plotted with regard to the frequency of the current.